\documentclass[twocolumn,preprintnumbers,amsmath,amssymb]{revtex4-1}
\usepackage{epsfig}
\usepackage{amsmath}
\usepackage{MnSymbol}
\usepackage{siunitx}
\usepackage{soul}
\usepackage{array}
\usepackage{multirow}
\usepackage{psfrag}
\usepackage[font=footnotesize,format=plain,labelsep=period,justification=raggedright]{caption}
\usepackage{subcaption}
\usepackage{xcolor}
\usepackage{graphicx}

\begin{document}
\title{Bidirectional Motion of Droplets on Liquid Infused Surfaces}
\author{Muhammad Subkhi Sadullah\textit{$^{a}$}, Gaby Launay\textit{$^{b}$}, Jayne Parle\textit{$^{a}$}, Rodrigo Ledesma-Aguilar\textit{$^{b}$}, Yonas Gizaw\textit{$^{c}$}, Glen McHale\textit{$^{b}$}, Gary Wells\textit{$^{b}$} and Halim Kusumaatmaja$^{\ast}$\textit{$^{a}$}}
\affiliation{$^{a}$Department of Physics, Durham University, Durham, DH1 3LE, UK.}
\affiliation{$^{b}$Smart Materials and Surfaces Laboratory, Northumbria University, Newcastle upon Tyne NE1 8ST, UK.}
\affiliation{$^{c}$The Procter and Gamble Co., Mason Business Center, 8700 S. Mason-Montgomery Road, Mason, OH, USA.}
\affiliation{$^{\ast}$Email: halim.kusumaatmaja@durham.ac.uk}

\begin{abstract}
	We demonstrate spontaneous bidirectional motion of droplets on liquid infused surfaces in the presence of a topographical gradient, in which the droplets can move either toward the denser or the sparser solid fraction area. Our analytical theory explains the origin of this bidirectional motion. Furthermore, using both lattice Boltzmann simulations and experiments, we show that the key factor determining the direction of motion is the wettability difference of the droplet on the solid surface and on the lubricant film. The bidirectional motion is shown for various combinations of droplets and lubricants, as well as for different forms of topographical gradients.
\end{abstract}

\maketitle
\renewcommand{\arraystretch}{1.5}

\section{Introduction}
Controlling droplet motion on a solid surface is important for a wide range of applications, from droplet microfluidics to water harvesting and self-cleaning surfaces \cite{Cho_2003, Mercedes_2007, Willmott_2011, Damak2018, Cammile_2019, Sun_2019}. Among the various approaches to induce motion, a good passive strategy is to introduce a wetting gradient on the solid surface, as this does not require energy to be provided continuously to the system. Such spontaneous motion has been extensively investigated for binary fluids systems under a variety of wetting gradients, including due to variations in surface chemistry \cite{Chaudhury_1539, Varnik2008}, topography \cite{Reyssat_2009,Moradi_2010,Lie_2016} and elasticity \cite{Style2013}.

More recently, there has been a growing interest to study droplet self-propulsion on liquid infused surfaces \cite{Zhang2018,McCarthy2019,Launay2019}. These are composite substrates constructed by infusing rough, textured or porous materials with wetting lubricants \cite{Wong2011,Lafuma_2011,Smith_2013}, which are known for their `slippery' properties. They have also been shown to exhibit a number of other advantageous surface properties, including anti-biofouling, anti-icing and self-healing \cite{Paxton_2017,Weisensee2017, Villegas_2019}.

\begin{figure}[h]
	\includegraphics[keepaspectratio]{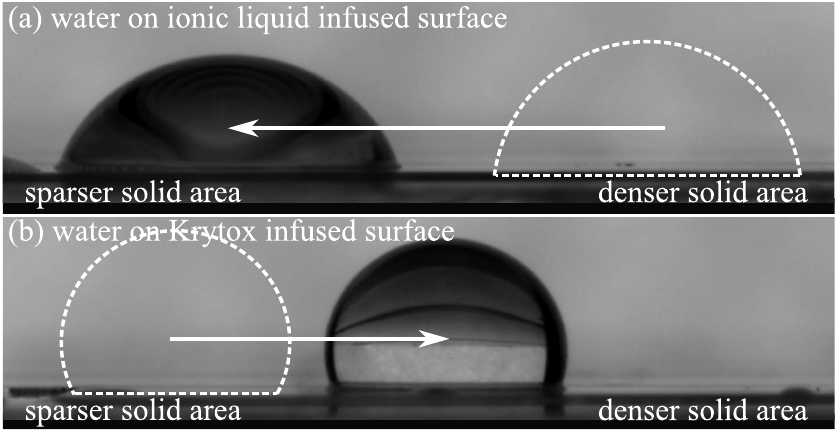}
	\caption{
		Spontaneous droplet motion on liquid infused surfaces with topographical gradient. (a) Water droplet on ionic liquid infused surface moves toward sparser solid area, while for (b) Krytox infused surface, water droplet moves toward denser solid area.
	}
	\label{fig:bidirectional_motion}
\end{figure}

Importantly, in all cases reported to date, including existing works on liquid infused surfaces, droplet motion on surfaces with texture/topographical gradients is always uni-directional towards the denser solid fraction area. In contrast, here we will demonstrate a bidirectional droplet motion. The presence of the lubricant on liquid infused surfaces can be exploited for a novel self-propulsion mechanism, in which the droplet has preferential wetting on {\em either} the denser {\em or} the sparser solid fraction area. Fig.~\ref{fig:bidirectional_motion} provides an example of this phenomenon. In Fig.~\ref{fig:bidirectional_motion}(a), when a structured substrate is infused with an ionic liquid, a water droplet placed on the surface moves toward the sparser solid area. In contrast, when the same substrate is infused with Krytox oil, the water droplet moves toward the denser solid area, as shown in Fig.~\ref{fig:bidirectional_motion}(b).

We structure our contribution as follows. First, we develop an analytical theory that elaborates how topographical gradient gives rise to the driving force that can propel droplets toward two possible directions. The spontaneous bidirectional motion depends on the combination of the solid, lubricant and droplet liquid and can occur as long as the lubricant does not fully wet the solid both in presence of the gas and the liquid droplet surroundings. We then verify our theory using both lattice Boltzmann simulations and experiments. We demonstrate this phenomenon can be observed using various liquid combinations for droplets and lubricants, as well as for different forms of structural gradients.

\section{Methods}
{\textbf{Numerical method.}}
Our numerical simulations are carried out employing a ternary free energy lattice Boltzmann method suitable for studying three fluids systems in complex geometries \cite{Semprebon2016a, Sadullah_2018}. The free energy model is given by
\begin{eqnarray}
\Psi & = & \sum_{m=1}^3  \int_\Omega \left(\frac{\kappa_m}{2} C_m^2 (1 - C_m)^2+ \frac{\alpha^2 \kappa_m}{2} (\nabla C_m)^2 \right) \, \mathrm{d}V \nonumber \\
&& - \sum_{m=1}^3  \int_{\partial\Omega} h_m C_m \, \mathrm{d}S,
\label{eq:ternary_free_energy}
\end{eqnarray}
where $C_m$ is the concentration of fluid phase $m$. In our simulations, $m=1,2,3$ represent the droplet, gas and lubricant phases respectively. The simulation parameters $\alpha$ and $\kappa$ are used to tune the interface width and surface tension, respectively. The $h_m$ parameters are related to the intrinsic contact angles of the fluids with the solid. The ESI provides additional details on how these parameters are chosen.

In the following, we set the local fluid density to be uniform, i.e., $\rho = C_1 + C_2 + C_3 =1$, since we expect that the effect of inertia is negligible for the droplet motion. Alternative simulation schemes are available for situations where the density difference between the fluid phases is important \cite{Moritz_2018,Neeru_2019}. Then, introducing the order parameters $\phi \equiv C_1 - C_2$, and $\psi \equiv C_3$ leads to the continuity, Navier-Stokes, and two Cahn-Hilliard equations
\begin{eqnarray}
& \partial_t \rho + \vec{\nabla} \cdot \left( \rho \vec{v} \right) = 0, \label{eq:continuity}  \\
& \partial_t (\rho \vec{v}) + \vec{\nabla} \cdot \left( \rho \vec{v}
\otimes \vec{v} \right) = - \vec{\nabla} \cdot \mathbf{P} + \vec{\nabla} \cdot
\left[ \eta ( \vec{\nabla} \vec{v} +  \vec{\nabla} \vec{v}^T ) \right], \,\,\,\,\,\,\, \label{eq:NSE}  \\
& \partial_t  \phi + \vec{\nabla} \cdot (\phi \vec{v}) = M_\phi \nabla^2 \mu_\phi, \label{eq:CHEphi} \\
&\partial_t \psi + \vec{\nabla} \cdot (\psi \vec{v}) = M_\psi \nabla^2 \mu_\psi, \label{eq:CHEpsi}
\end{eqnarray}
where $\vec{v}$ and $\eta$ are the fluid velocity and viscosity respectively. 
Eqs.~\eqref{eq:CHEphi} and \eqref{eq:CHEpsi} describe the evolution of $\phi$ and $\psi$, and, correspondingly, the interfaces between the three fluids. The thermodynamic properties of the system, described in the free energy model in Eq. (\ref{eq:ternary_free_energy}), enter the equations of motion via the chemical potentials, $\mu_q = \delta \Psi/ \delta q$, ($q = \phi$ and $\psi$), and the pressure tensor, $\bf{P}$, defined by $\partial_\beta P_{\alpha\beta} = \phi\partial_\alpha \mu_\phi + \psi\partial_\alpha \mu_\psi$. The equations of motion in Eqs.~\eqref{eq:continuity}-\eqref{eq:CHEpsi} are solved using the lattice Boltzmann method \cite{Brient2004,Semprebon2016a}.

{\textbf{Experimental method.}}
For the experiments, we use photolithography to produce surfaces with \SI{60}{\um} deep grooves in the $x$-direction. The width of each groove can be tuned (between $10$ and \SI{75}{\um}) to obtain solid fractions $f_s$ ranging from $0.1$ to $0.9$.
This allows us to create topographical gradients along the $x$-direction by continuously increasing or decreasing the width of the grooves.
After fabrication, the geometry of the surfaces is carefully measured using optical profilometry and SEM (Scanning Electron Microscope) imaging.

To reduce the contact angle hysteresis that would hinder droplet motion, the structured surfaces are treated with SOCAL (Slippery Omniphobic Covalently-Attached Liquid), following the protocol from Wang \textit{et al.} \cite{Wang_2015}, modified for SU-8 substrates (see ESI for details).
We verify the SOCAL coating by measuring the contact angle ($\SI{104.2}{\degree} \pm \SI{2}{\degree}$) and contact angle hysteresis ($<\SI{5}{\degree}$) of a water droplet deposited on a non-structured (flat) region of the sample.

The surfaces are then dipped in a lubricant and left to drain vertically for \SI{10}{\minute}, in order to fill the grooves and create a liquid infused surface.
Droplets are finally deposited on the imbibed surfaces using a thin needle and their motion is tracked using a camera placed on the side.
To rule out the effect of gravity on the droplet motion, the surface is slightly tilted ($\approx\SI{0.5}{\degree}$) against the direction of motion. The procedure is repeated $5$ times for each configuration to ensure reproducibility. The sample fabrication details are further elaborated in the ESI.

\section{Results and Discussions}
{\textbf{The origin of the driving force.}}
When a liquid droplet is placed on a homogenous solid surface, it stays stationary because the surface tension force pulls the base of the droplet equally in the radial direction \cite{Young_1805}. This force balance is broken when the wettability of one side of the droplet is different from the other, resulting in a spontaneous droplet motion towards the more wettable region of the solid \cite{Subramanian_2005}.

On liquid infused surfaces, the apparent contact angle of a droplet depends on the surface tensions and the intrinsic contact angles of all fluids involved in the system \cite{Semprebon2016b,Kreder_2018, McHale_2019}. This rich interplay makes it much less trivial to predict the direction of droplet motion when there is a topographical gradient. To do this we need to break down the contributing surface tension forces.

\begin{figure}
	\centering
	\includegraphics[keepaspectratio]{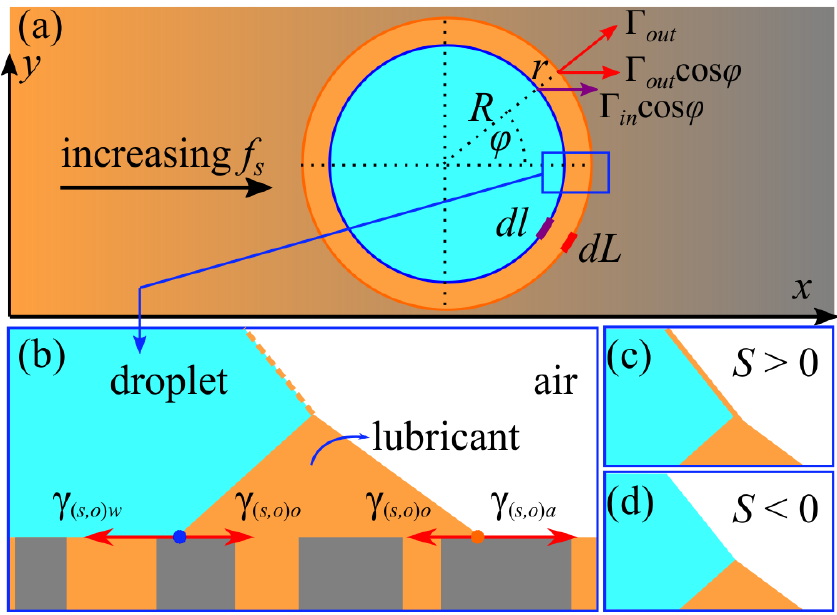}
	\caption{
		(a) Illustration of a droplet on a liquid infused surface with topographical gradient, where greater solid fraction ($f_s$) area is indicated by the darker area. $R$ and $r$ are the droplet base radius and meniscus width respectively. $\Gamma_{out}$ and $\Gamma_{in}$ are the surface tension forces per unit length that act on the outer and inner contact lines. (b) Magnification of the meniscus area (side-view). The red arrows indicates the relevant composite interfacial tensions, as described in Eqs.~\eqref{eq:gamma_eff1}-\eqref{eq:gamma_eff3}. The dashed line at droplet-air interface indicates the possibility of lubricant encapsulation. Depending on the sign of the spreading parameter $S$, the lubricant may encapsulate the droplet (c-d).
	}
	\label{fig:illustration}
\end{figure}

Consider a liquid droplet placed on top of a liquid infused surface with topographical gradient, as shown in Fig.~\ref{fig:illustration}(a). The substrate is set horizontally such that gravity does not play a role. For convenience, we use the subscripts $w,o,a$ and $s$ to refer to the droplet, infusing lubricant, air and solid phases respectively. Furthermore, we introduce the spreading parameter \cite{Smith_2013},
\begin{equation}
S=\gamma_{wa}-\gamma_{oa}-\gamma_{ow},
\label{eq:spreading}
\end{equation}
with $\gamma_{\alpha\beta}$ the interfacial tension between phases $\alpha$ and $\beta$. The droplet is encapsulated by the lubricant when $S>0$ \cite{Smith_2013,Daniel2017,Kreder_2018}, see Fig.~\ref{fig:illustration}(c). For $S<0$, the droplet is not encapsulated, as illustrated in Fig.~\ref{fig:illustration}(d). 

We will now argue that liquid infused surfaces can be considered as composite surfaces of solid and lubricant, with fractions of $f_s$ and $(1-f_s)$ respectively. 
Therefore, the {\em composite} interfacial tension of the liquid infused surface with phase $\alpha$ is $\gamma_{(s,o)\alpha} \equiv f_s \gamma_{s\alpha}+(1-f_s)\gamma_{o\alpha}$.
Letting the solid fraction $f_s$ vary in the $x$ direction only leads to the interfacial tensions (Fig.~\ref{fig:illustration}(b))
\begin{align}
\gamma_{(s,o)w}&\equiv f_s(x)\gamma_{sw}+(1-f_s(x))\gamma_{ow},\label{eq:gamma_eff1}\\
\gamma_{(s,o)o}&\equiv f_s(x)\gamma_{so}+(1-f_s(x))\gamma_{oo}=f_s(x)\gamma_{so},\label{eq:gamma_eff2}\\
\gamma_{(s,o)a}&\equiv f_s(x)\gamma_{sa}+(1-f_s(x))\gamma_{oa}.\label{eq:gamma_eff3}
\end{align}

The relevant surface tension forces per unit length that pull the droplet in radial direction are $\Gamma_{in}=\gamma_{(s,o)o}-\gamma_{(s,o)w}$ and $\Gamma_{out}=\gamma_{(s,o)a}-\gamma_{(s,o)o}$ for the inner (droplet-lubricant-composite substrate) and the outer (lubricant-air-composite substrate) contact lines respectively. As detailed in the ESI, we assume that the drop shape is in quasi-equilibrium, so that the net contributions from the droplet-air, droplet-lubricant and lubricant-air surface tensions go to zero. Furthermore, since $f_s$ does not vary with $y$, only the $x$-component of the forces contributes to the driving force, i.e. $\Gamma_{in}\cos{\varphi}$ and $\Gamma_{out}\cos{\varphi}$ (see Fig.~\ref{fig:illustration}(a)). The total driving force is thus the sum of these surface tensions integrated over the total perimeters of the inner and outer contact lines,
\begin{equation}
F=\int_{l} \Gamma_{in}\cos{\varphi} dl + \int_{L} \Gamma_{out}\cos{\varphi} dL.
\label{eq:FD_full}
\end{equation}
Assuming the droplet base is circular, we can express $dl = Rd\varphi$ and $dL = (R+r)d\varphi$. Moreover, if the meniscus is much smaller than the droplet base radius, we can approximate $R+r \approx R$, and thus, $dl=dL=Rd\varphi$.
The finite meniscus size case is described in the ESI.

\begin{figure*}[t!]
	\centering
	\includegraphics[keepaspectratio]{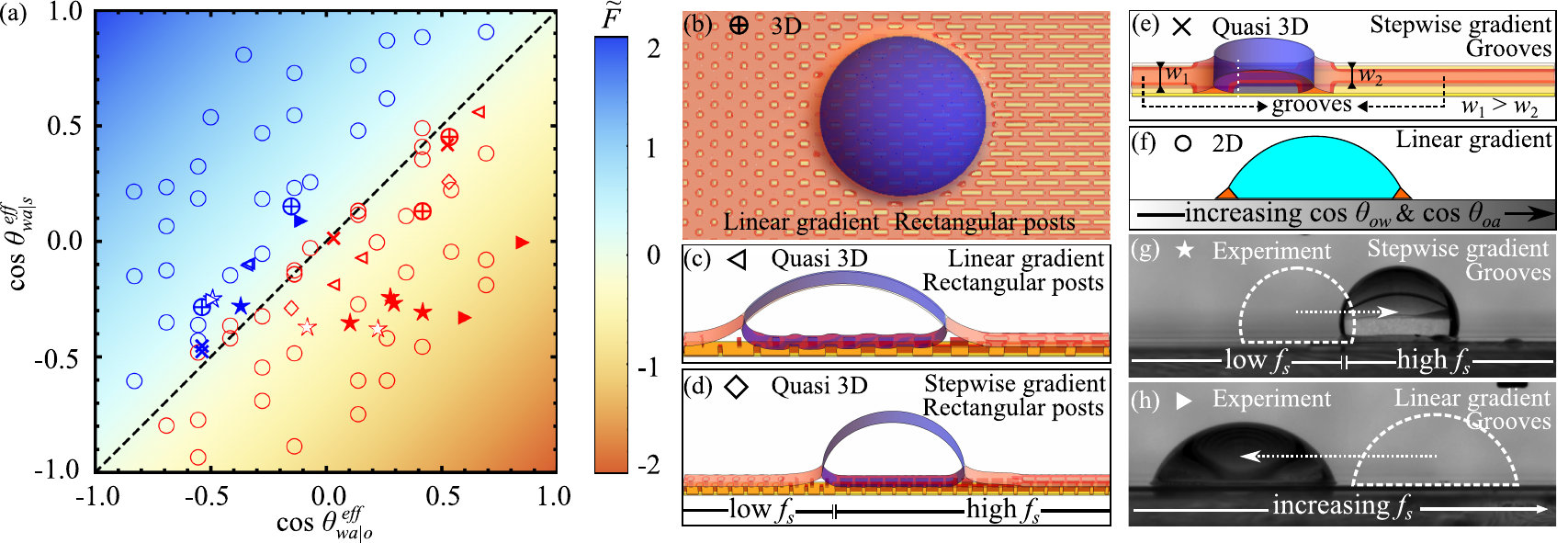}
	\caption{
		(a) Confirmation of the bidirectional motion of droplets on liquid infused surfaces with a topographical gradient, as predicted in Eq.~(\ref{eq:FDN}) via simulations and experiments. The blue and red data points indicate that the droplets were observed to move to the higher or lower solid fraction area, which respectively correspond to possitive and negative $\tilde{F}$. The symbols in the phase diagram correspond to the simulation and experimental setups explained in panels (b-h). For (g) and (h), the hollow $\largestar$ and $\triangleright$ data points indicate the lubricant encapsulation case.
	}
	\label{fig:phase_diagram_and_experiments}
\end{figure*}

In this vanishing meniscus approximation, we can substitute the definitions of the composite interfacial tensions in Eqs.~\eqref{eq:gamma_eff1}-\eqref{eq:gamma_eff3} to Eq.~\eqref{eq:FD_full}, and write the driving force as
\begin{align}
F=& \big((\gamma_{so}-\gamma_{sw}+\gamma_{ow})+(\gamma_{sa}-\gamma_{so}-\gamma_{oa})\big) \, \times \nonumber\\
&\int_{\varphi}f_s(x)R\cos{\varphi} d\varphi.\label{eq:FD_first1}
\end{align}
We can simplify Eq.~\eqref{eq:FD_first1} by employing the Young's contact angles of the lubricant in the air and in the droplet phase environment, respectively defined as $\cos{\theta_{oa}}=(\gamma_{sa}-\gamma_{so})/\gamma_{oa}$ and $\cos{\theta_{ow}}=(\gamma_{sw}-\gamma_{so})/\gamma_{ow}$. In this case, Eq.~\eqref{eq:FD_first1} becomes
\begin{align}
F= &\big(\gamma_{ow}(1-\cos{\theta_{ow}})+\gamma_{oa}(\cos{\theta_{oa}}-1)\big) \, \times \nonumber \\
&\int_{\varphi}f_s(x)R\cos{\varphi} d\varphi.\label{eq:interpret1}
\end{align}
We find that the driving force ceases ($F = 0$) only if the lubricant completely wets the solid surface both in the air and in the droplet phase environments, such that $\theta_{ow} = \theta_{oa} = 0$. 
This is expected since, in this case, the surface topography is covered by a thin layer of lubricant  everywhere. We can still expect spontaneous motion to occur if either $\theta_{ow}$ or $\theta_{oa}$ is non-zero.

To determine the direction of droplet motion, we can introduce the droplet-air effective interfacial tension \cite{McHale_2019}
\begin{equation}
\gamma_{eff}\equiv\begin{cases}
\gamma_{oa}+\gamma_{ow}, & \text{if } S>0 \text{  (lubricant encapsulation)},\\
\gamma_{wa}, & \text{otherwise,}
\end{cases}\nonumber
\end{equation}
and the following definitions of apparent contact angles
\begin{equation}
\cos{\theta_{wa|s}^{eff}} \equiv\frac{\gamma_{sa}-\gamma_{sw}}{\gamma_{eff}}, \quad\cos{\theta_{wa|o}^{eff}} \equiv\frac{\gamma_{oa}-\gamma_{ow}}{\gamma_{eff}},
\label{eq:apparent_contact_angles}
\end{equation}
such that the driving force in Eq.~\eqref{eq:FD_first1} can be written in the following form
\begin{equation}
F=\gamma_{eff}\big(\cos{\theta_{wa|s}^{eff}}-\cos{\theta_{wa|o}^{eff}}\big)\int_{\varphi} f_s(x)\cos{\varphi} R d\varphi.
\label{eq:FD}
\end{equation}
$\theta_{wa|s}^{eff}$ and $\theta_{wa|o}^{eff}$ are defined as the contact angles of the droplet, either encapsulated by lubricant or not, on a smooth solid surface and on the lubricant surface respectively. When there is no encapsulation, $\gamma_{eff}=\gamma_{wa}$ and hence $\theta_{wa|s}^{eff}=\theta_{wa|s}$, which is the familiar Young's contact angle of a droplet on a smooth solid surface \cite{Young_1805}. 

Let us now discuss the terms in Eq.~\eqref{eq:FD}. The term under the integral depends on the details of the surface patterning, $f_s(x)$, and it modulates the strength of the driving force.
The direction of the driving force is determined only by the sign of the gradient in $f_s(x)$ and by the prefactor
\begin{equation}
\tilde{F}=\big(\cos{\theta_{wa|s}^{eff}}-\cos{\theta_{wa|o}^{eff}}\big),
\label{eq:FDN}
\end{equation}
which is in fact independent of the surface texture.
This has a clear and intuitive physical interpretation: it corresponds to the preferential wetting of the droplet on the region exhibiting the majority of solid or lubricant surface. Without any loss of generality, let us assume that the gradient in $f_s(x)$ is positive, i.e. the solid fraction becomes denser with increasing $x$. When $\cos{\theta_{wa|s}^{eff}}>\cos{\theta_{wa|o}^{eff}}$, the droplet prefers to wet the solid rather than the lubricant. Therefore, the droplet moves toward the solid majority surface (denser solid area). In contrast, when $\cos{\theta_{wa|s}^{eff}}<\cos{\theta_{wa|o}^{eff}}$, the droplet moves toward lubricant majority surface (sparser solid area).

{\textbf{Demonstration of Bidirectional Motion using Simulations and Experiments.}}
To validate the prediction of Eq.~(\ref{eq:FDN}), we perform both simulations and experiments of droplets moving across liquid infused surfaces with textural gradients. The details of the simulation and experimental methods are provided in the Method section and in the ESI. 

Fig.~\ref{fig:phase_diagram_and_experiments} shows a phase diagram for the normalised driving force ($\tilde F$), predicted by Eq~(\ref{eq:FDN}) (colormap), and the corresponding droplet motion observed in the numerical simulations and the experiments (symbols). The upper section of the phase map corresponds to an expected driving force directed towards the denser solid regions, while the lower section towards the sparser solid regions. The color of the symbols represents motion to the denser (blue) or sparser (red) solid fraction area, showing a good agreement between the numerical simulations and the experiments with the theoretical prediction. 

Our numerical simulations show that the mechanism leading to bidirectional motion holds for different surface topographies, and thus supports that the relevant control parameter linked to the topography of the solid is the solid fraction $f_s$. Specifically, we consider three different simulation geometries. Firstly, we use full 3D simulations with linear gradient of rectangular posts ($\oplus$, Fig.~\ref{fig:phase_diagram_and_experiments}(b)). For the linear gradient, the post length is increased for each subsequent post in the $x$-direction. Secondly, we carry out quasi 3D simulations, where a cylindrical droplet and only a period of the surface features in the $y$ direction are used. Here we employ both a linear gradient of rectangular posts ($\triangleleft$, Fig.~\ref{fig:phase_diagram_and_experiments}(c)), as well as stepwise gradients of rectangular posts ($\Diamond$, Fig.~\ref{fig:phase_diagram_and_experiments}(d)) and grooves ($\times$, Fig.~\ref{fig:phase_diagram_and_experiments}(e)). In the case of a stepwise gradient, the substrate is divided into lower and higher $f_s$ regimes. Thirdly, we use 2D simulations ($\rm{\bigcircle}$, Fig.~\ref{fig:phase_diagram_and_experiments}(f)). Here, the topographical gradient is not simulated explicitly, but instead it is represented by varying the effective lubricant-droplet contact angle $\theta_{ow}(x)$ and the effective lubricant-air contact angle $\theta_{oa}(x)$ \cite{Cassie_Baxter_1944}:
\begin{equation}
\cos{\theta_{o\alpha}}(x)= f_s(x) \cos{\theta_{o\alpha}^{\rm Y}} + (1 - f_s(x)),
\end{equation}
where the subscript $\alpha = w, a$ and $\cos{\theta_{o\alpha}^{\rm Y}}$ is the contact angle on the smooth flat surface. In Fig.~\ref{fig:phase_diagram_and_experiments}, few exceptions are present for the 2D simulations, where some of the red data points cross the diagonal line in the phase diagram. This is due to the finite size effect of the lubricant meniscus. As explained in the ESI, such finite size effect becomes relevant for $\tilde{F}\approx0$ (close to the diagonal line in the phase diagram).
 
Our experimental results correspond to two different solid surface geometries: stepwise ({\LARGE $\star$}) and linear ({\large $\blacktriangleright$}) gradients (see Fig.~\ref{fig:phase_diagram_and_experiments}(g-h)); and, crucially, show that the direction of motion of a droplet on a given topography can be switched by choosing the interfacial tensions. In Fig.~\ref{fig:phase_diagram_and_experiments} we report experimental results for water droplets in contact with ten different lubricants and ethylene glycol droplets in contact with two different lubricants. In the phase diagram, the hollow and filled symbols correspond to cases where the droplet is encapsulated and not encapsulated by the lubricant, respectively.

\begin{figure}
	\centering
	\includegraphics[width=\linewidth, keepaspectratio]{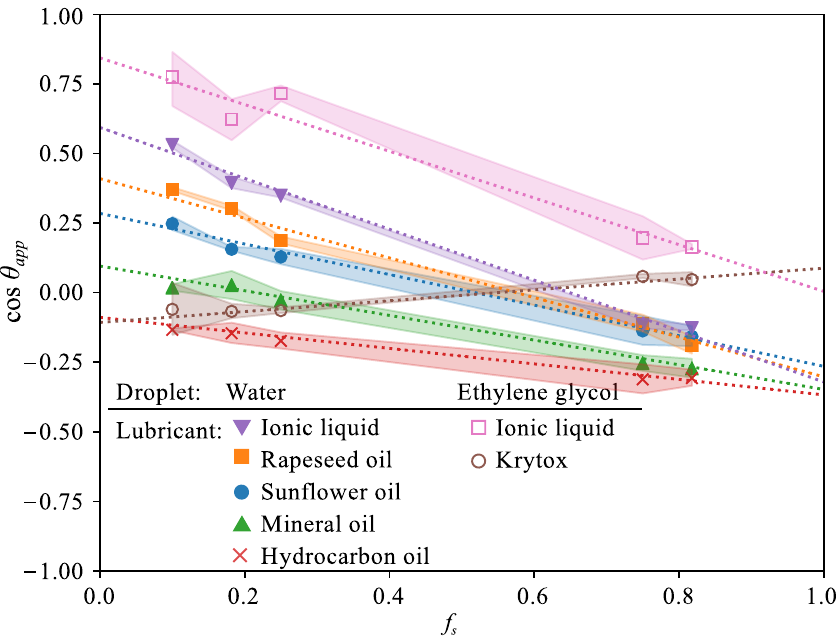}
	\caption{
		Estimation of normalised driving force $\tilde{F}$ for water and ethylene glycol droplets on structured surfaces imbibed with various lubricants.
		Each point is the average of $5$ contact angle measurements of sessile droplets.
		The surrounding coloured area represents the standard deviation.
		Dashed lines are fits of Eq.~(\ref{eq:fit_graph_method}) using the least-square method. The gradient of the fits corresponds to $\tilde{F}$,
		while extrapolations of those fits to $f_s=1$ and $f_s=0$ give a measure of the values of $\cos\theta_{wa|s}^{eff}$ and $\cos\theta_{wa|o}^{eff}$ respectively.
	}
	\label{fig:graphical_method}
\end{figure}

To position the experimental data points in the phase diagram, it is necessary to infer the effective wettability of the surface, given by $\cos{\theta_{wa|s}^{eff}}$ and $\cos{\theta_{wa|o}^{eff}}$. If the values of $\theta_{wa|s}, \gamma_{wa}, \gamma_{oa}$ and $\gamma_{ow}$ are known in the literature \cite{Girifalco1957}, they can simply be calculated from Eq.~\eqref{eq:apparent_contact_angles}. We are able to calculate these for five different droplet-lubricant combinations, as tabulated in the ESI. Alternatively, we can determine $\cos{\theta_{wa|s}^{eff}}$ and $\cos{\theta_{wa|o}^{eff}}$ using a graphical method as follows. In the vanishing meniscus approximation, the droplet apparent contact angle on the composite solid-lubricant surface can be expressed as \cite{Semprebon2016b,McHale_2019}
\begin{align}
\cos{\theta_{app}}&=\frac{\gamma_{(s,o)a}-\gamma_{(s,o)w}}{\gamma_{eff}},\\
&=\big(\cos{\theta_{wa|s}^{eff}}-\cos{\theta_{wa|o}^{eff}}\big)f_s+\cos{\theta_{wa|o}^{eff}},\label{eq:fit_graph_method}\\
&=\tilde{F}f_s+\cos{\theta_{wa|o}^{eff}}.\label{eq:fit_graph_method2}
\end{align}
As shown in Fig. \ref{fig:graphical_method} for seven separate droplet-lubricant pairs, by measuring $\theta_{app}$ for different values of the solid fraction $f_s$, we can determine the normalised driving force $\tilde{F}$ from the gradient of the curve. Furthermore, $\cos{\theta_{wa|s}^{eff}}$ and $\cos{\theta_{wa|o}^{eff}}$ can be inferred by extrapolating the curve to $f_s = 1$ and $f_s = 0$. All experimental values of $\cos{\theta_{wa|s}^{eff}}$, $\cos{\theta_{wa|o}^{eff}}$ and consequently $\tilde{F}$ used in Fig.~\ref{fig:phase_diagram_and_experiments} are provided in the ESI.

\section{Conclusions}
We have reported a spontaneous bidirectional motion of droplet on liquid infused surfaces with topographical gradient. In contrast to previous studies describing uni-directional droplet motion on surfaces with topographical gradients, here the droplet can move toward the sparser or the denser solid fraction area. We investigated the origin of this bidirectional motion by looking into the relevant surface tension forces acting on the droplet. Our analytical theory predicts, and our simulation and experimental results confirmed, that the direction of the motion is determined by a simple physical quantity, $(\cos{\theta_{wa|s}^{eff}}-\cos{\theta_{wa|o}^{eff}})$. This quantity can be intuitively interpreted as preferential wetting of the droplet on the solid majority surface (denser solid area) or on the lubricant majority surface (sparser solid area). The bidirectional motion is also validated over a wide range of surface tension and contact angle combinations, with and without lubricant encapsulation, and for different types of topographical gradients, both in our simulations and experiments.

There are a number of avenues of future work to better understand and exploit the novel phenomenon described here. For instance, while we already show here that bidirectional motion applies for different types of topographical gradients, it remains an open problem which types of topographical gradients are optimal. It is also an interesting to study the detailed dynamics of the droplets under wetting gradients, including how the droplet velocity can be systematically controlled. Moreover, since different droplet-lubricant combination may move to different direction, we envisage it can be exploited to sort droplets based on their interfacial property; and when combined with gravity, simultaneously based on their size and interfacial property, by playing off the competition between the forces due to wetting gradient and due to gravity. More complex applications include liquid/liquid separation or directing chemical reactions in a droplet microfluidic device.

\section{Acknowledgements}
M.S.S. is supported by an LPDP (Lembaga Pengelola Dana Pendidikan) scholarship from the Indonesian Government. H.K. acknowledges funding from EPSRC (grant EP/P007139/1) and Procter and Gamble. G.G.W. and G.L. acknowledge funding from EPSRC (grant EP/P026613/1).


\bibliographystyle{rsc} 
\bibliography{template.article}


\end{document}